\documentclass[11pt]{article}
\usepackage{epsfig}

\begin{document}

\title{{\Large\bf Quantum spacetime: what do we know?}}

\author{{Carlo Rovelli}\\
{Centre de Physique Theorique de Luminy, Marseille, 
France}\\  
{Physics Department, University of Pittsburgh, Pittsburgh, USA}}

\date{March 8th, 1999}

\maketitle

\centerline{\em To appear on ``Physics Meets Philosophy at the Planck 
scale''}
\centerline{\em
C Callender N Hugget eds, Cambridge University 
Press}

\begin{abstract}

I discuss nature and origin of the problem of quantum gravity.  I 
examine the knowledge that may guide us in addressing this 
problem, and the reliability of such knowledge.  In particular, I 
discuss the subtle modification of the notions of space 
and time engendered by general relativity, and how these might 
merge into quantum theory.  I also present some reflections on 
methodological questions, and on some general issues in 
philosophy of science which are are raised by, or a relevant for, 
the research on quantum gravity.

\end{abstract}

\section{The incomplete revolution}

Quantum mechanics (QM) and general relativity (GR) have modified 
our understanding of the physical world in depth.  But they have 
left us with a general picture of the physical world which is 
unclear, incomplete, and fragmented.  Combining what we have 
learn about our world from the two theories and finding a new 
synthesis is a major challenge, perhaps {\em the\/} major 
challenge, in today's fundamental physics.

The two theories have a opened a major scientific revolution, but 
this revolution is not completed.  Most of the physics of this 
century has been a sequel of triumphant explorations of the new 
worlds opened by QM and GR. QM lead to nuclear physics, solid 
state physics, and particle physics.  GR to relativistic 
astrophysics, cosmology and is today leading us towards 
gravitational astronomy.  The urgency of applying the two 
theories to larger and larger domains, the momentous 
developments, and the dominant pragmatic attitude of the middle 
of this century, have obscured the fact that a consistent picture 
of the physical world, more or less stable for three centuries, 
has been lost with the advent of QM and GR. This pragmatic 
attitude cannot be satisfactory, or productive, in the long run.  
The basic Cartesian-Newtonian notions such as matter, space, 
time, causality, have been modified in depth.  The new notions do 
not stay together.  At the basis of our understanding of the 
world reigns a surprising confusion.  From QM and GR we know that 
we live in a spacetime with quantum properties, that is, a {\em 
quantum spacetime}.  But what is a quantum spacetime?

In the last decade, the attention of the theoretical physicists 
has been increasingly focusing on this major problem.  Whatever 
the outcome of the enterprise, we are witnessing a large scale 
intellectual effort for accomplishing a major aim: completing the 
XXth scientific revolution, and finding a new synthesis.

In this effort, physics is once more facing {\em conceptual\/} 
problems: {\em What is matter?  What is causality?  What is the 
role of the observer in physics?  What is time?  What is the 
meaning of ``being somewhere''?  What is the meaning of ``now''?  
What is the meaning of ``moving''?  Is motion to be defined with 
respect to objects or with respect to space?\/} These 
foundational questions, or sophisticated versions of these 
questions, were central in the thinking and in the results of 
Einstein, Heisenberg, Bohr, Dirac and their colleagues.  But 
these are also precisely the same questions that Descartes, 
Galileo, Huygens, Newton and their contemporaries debated with 
passion -- the questions that lead them to create modern science.  
For the physicists of the middle of this century, these questions 
were irrelevant: one does not need to worry about first 
principles in order to apply the Schr\"odinger equation to the 
helium atom, or to understand how a neutron star stays together.  
But today, if we want to find a novel picture of the world, if we 
want to understand what is quantum spacetime, we have to return, 
once again, to those foundational issues.  We have to find a new 
answer to these questions --different from Newton's answer-- 
which took into account what we have learned about the world with 
QM and GR. 

Of course, we have little, if any, direct empirical access to the 
regimes in which we expect genuine quantum gravitational 
phenomena to appear.  Anything could happen at those 
fantastically small distance scales, far removed from our 
experience.  Nevertheless, we do have information about quantum 
gravity, and we do have indications on how to search it.  In 
fact, we are precisely in one of the very typical situations in 
which good fundamental theoretical physics has been working at 
its best in the past: we have learned two new extremely general 
``facts'' about our world, QM and GR, and we have ``just'' to 
figure out what they imply, when taken together.  The most 
striking advances in theoretical physics happened in situations 
analogous to this one.

Here, I present some reflections on these issues.\footnote{For 
recent general overviews of current approaches to quantum gravity, see 
(Isham 1999) and (Rovelli 1999).}  What have we 
learned about the world from QM and, especially, GR? What do we 
know about space, time and matter?  What can we expect from a 
quantum theory of spacetime?  To which extent does taking QM and 
GR into account force us to modify the notion of time?  What can 
we already say about quantum spacetime?

I present also a few reflections on issues raised by the relation 
between philosophy of science and research in quantum gravity.  I 
am not a philosopher, and I can touch philosophical issues only 
at the risk of being naive.  I nevertheless take this risk here, 
encouraged by Craig Callender and Nick Huggett extremely 
stimulating idea of this volume.  I present some methodological 
considerations --How shall we search?  How can the present 
successful theories can lead us towards a theory that does not 
yet exist?-- as well as some general consideration.  In 
particular, I discuss the relation between physical theories that 
supersed each others and the attitude we may have with respect to 
the truth-content of a physical theory, with respect to the 
reality of the theoretical objects the theory postulates in 
particular, and to its factual statements on the world in 
general.

I am convinced of the reciprocal usefulness of a dialog between 
physics and philosophy (Rovelli 1997a).  This dialog has played a 
major role during the other periods in which science faced 
foundational problems.  In my opinion, most physicists 
underestimate the effect of their own epistemological prejudices 
on their research.  And many philosophers underestimate the 
influence --positive or negative-- they have on fundamental 
reserach.  On the one hand, a more acute philosphical awarness 
would greatly help the physicists engaged in fundamental 
research: Newton, Heisenberg and Einstein couldn't have done what 
they have done if they weren't nurtured by (good or bad) 
philosophy.  On the other hand, I wish contemporary philosophers 
concerned with science would be more interested in the ardent 
lava of the foundational problems science is facing today.  It is 
here, I believe, that stimulating and vital issues lie.

\section{The problem} 

What is the task of a quantum theory of gravity, and how should 
we search for such a theory?  The task of the search is clear and 
well defined.  It is determined by recalling the three major  
steps that lead to the present problematic situation.

\subsection{First step. A new actor on the stage: the field}

The first step is in the works of Faraday, Maxwell and Einstein.  
Faraday and Maxwell have introduced a new fundamental notion in 
physics, the field.  Faraday's book includes a fascinating 
chapter with the discussion of whether the field (in Faraday's 
terminology, the ``lines of force'') is ``real''.  As far as I 
understand this subtle chapter (understanding Faraday is tricky: 
it took the genius of Maxwell), in modern terms what Faraday is 
asking is whether there are independent degrees of freedom in the 
electric and magnetic fields.  A degree of freedom is a quantity 
that I need to specify (more precisely: whose value and whose 
time derivative I need to specify) in order to be able to predict 
univocally the future evolution of the system.  Thus Faraday is 
asking: if we have a system of interacting charges, and we know 
their positions and velocities, is this knowledge sufficient to 
predict the future motions of the charges?  Or rather, in order 
to predict the future, we have to specify the instantaneous 
configuration of the field (the fields degrees of freedom), as 
well?  The answer is in Maxwell equations: the field has 
independent degrees of freedom.  We cannot predict the future 
evolution of the system from its present state unless we know the 
instantaneous field configuration.  Learning to use these degrees 
of freedom lead to radio, TV and cellular phone.

To which physical entity do the degrees of freedom of the 
electromagnetic field refer?  This was one of the most debated 
issues in physics towards the end of last century.  The 
electromagnetic waves have aspects in common with water waves, or 
with sound waves, which describe vibrations of some material 
medium.  The natural interpretation of the electromagnetic field 
was that it too describes the vibrations of some material medium 
-- for which the name ``ether'' was chosen.  A strong argument  
supports this idea: The wave equations for water or sound waves 
fail to be Galilean invariant.  They do so because they describe 
propagation over a medium (water, air) whose state of motion 
breaks Galilean invariance and defines a preferred reference 
frame.  Maxwell equations break Galilean invariance as well and 
it was thus natural to hypothesize a material medium determining 
the preferred reference frame.  But a convincing dynamical theory 
of the ether compatible with the various experiments (for 
instance on the constancy of the speed of light) could not be 
found.

Rather, physics took a different course.  Einstein {\em 
believed\/} Maxwell theory as a fundamental theory and {\em 
believed\/} the Galilean insight that velocity is relative and 
inertial system are equivalent.  Merging the two, he found 
special relativity.  A main result of special relativity is that 
the field cannot be regarded as describing vibrations of 
underlying matter.  The idea of the ether is abandoned, and {\em 
the field has to be taken seriously as elementary constituent of 
reality}.  This is a major change from the ontology of 
Cartesian-Newtonian physics.  In the best description we can give 
of the physical world, there is a new actor: the field.  The 
electromagnetic field can be described by the Maxwell potential 
$A_{\mu}(x),\ \mu=0,1,2,3$.  The entity described by $A_{\mu}(x)$ 
(more precisely, by a gauge-equivalent class of $A_{\mu}(x)$'s) 
is one of the elementary constituents of the physical world, 
according to the best conceptual scheme physics has find, so far, 
for grasping our world. 

\subsection{Second step. Dynamical entities have quantum properties}

The second step (out of chronological order) is the replacement 
of the mechanics of Newton, Lagrange and Hamilton with quantum 
mechanics (QM).  As did classical mechanics, QM provides a very 
general framework.  By formulating a specific dynamical theory 
within this framework, one has a number of important physical 
consequences, substantially different from what is implied by the 
Newtonian scheme.  Evolution is probabilistically determined 
only; some physical quantities can take certain discrete values 
only (are ``quantized''); if a system can be in a state $A$, 
where a physical quantity $q$ has value $a$, as well as in state 
$B$, where $q$ has value $b$, then the system can also be in 
states (denoted $\Psi=c_{a}A+c_{b}B$) where $q$, has value $a$ 
with probability $|c_{a}|^{2}/(|c_{a}|^{2}+|c_{b}|^{2})$, or, 
alternatively, $b$ with probability $|c_{b}|^{2} /(|c_{a}|^{2} 
+|c_{b}|^{2})$ (superposition principle); conjugate variables 
cannot be assumed to have value at the same time (uncertainty 
principle); and what we can say about the properties that the 
system will have the-day-after-tomorrow is not determined just by 
what we can say about the system today, but also on what we will 
be able to say about the system tomorrow.  (Bohr would had simply 
said that observations affect the system.  Formulations such as 
Bohm's or consistent histories force us to use intricate wording 
for naming the same physical fact.)

The formalism of QM exists in a number of more or less equivalent 
versions: Hilbert spaces and self-adjoint observables, Feynman's 
sum over histories, algebraic formulation, and others.  Often, we 
are able to translate from one formulation to another.  However, 
often we cannot do easily in one formulation, what we can do in 
another.

QM is not the theory of micro-objects.  It is our best form of 
mechanics.  If quantum mechanics failed for macro-objects, we 
would have detected the boundary of its domain of validity in 
mesoscopic physics.  We haven't.\footnote{Following Roger 
Penrose's opposite suggestions of a failure of conventional QM 
induced by gravity (Penrose 1995), Antony Zeilinger is preparing 
an experiment to test such a possible failure of QM (Zeilinger 
1997).  It would be very exciting if Roger turned out to be 
right, but I am afraid that QM, as usual, will win.} The 
classical regime raises some problems (why effects of macroscopic 
superposition are difficult to detect?).  Solving these problems 
requires good understanding of physical decoherence and perhaps 
more.  But there is no reason to doubt that QM represents a 
deeper, not a shallower level of understanding of nature than 
classical mechanics.  Trying to resolve the difficulties in our 
grasping of our quantum world by resorting to old classical 
intuition is just lack of courage.  We have learned that the 
world has quantum properties.  This discovery will stay with us, 
like the discovery that velocity is only relational or like the 
discovery that the Earth is not the center of the universe.

The empirical success of QM is immense.  Its physical obscurity 
is undeniable.  Physicists do not yet agree on what QM precisely 
says about the world (the difficulty, of course, refers to 
physical meaning of notions such as ``measurement'', ``history'', 
``hidden variable'', \ldots).  It is a bit like the Lorentz 
transformations before Einstein: correct, but what do they mean?

In my opinion, what QM means is that the contingent (variable) 
properties of any physical system, or the state of the system, 
are relational notion which only make sense when referred to a 
second physical system.  I have argued for this thesis in 
(Rovelli 1996, Rovelli 1998).  However, I will not enter in this 
discussion here, because the issue of the interpretation of QM 
has no direct connection with quantum gravity.  Quantum gravity 
and the interpretation of QM are two major but (virtually) 
completely unrelated problems.

QM was first developed for systems with a finite number of 
degrees of freedom.  As discussed in the previous section, 
Faraday, Maxwell and Einstein had introduced the field, which has 
an infinite number of degrees of freedom.  Dirac put the two 
ideas together.  He {\em believed\/} quantum mechanics and he 
{\em believed\/} Maxwell's field theory much beyond their 
established domain of validity (respectively: the dynamics of 
finite dimensional systems, and the classical regime) and 
constructed quantum field theory (QFT), in its first two 
incarnations, the quantum theory of the electromagnetic field and 
the relativistic quantum theory of the electron.  In this 
exercise, Dirac derived the existence of the photon just from 
Maxwell theory and the basics of QM. Furthermore, by just {\em 
believing\/} special relativity and {\em believing\/} quantum 
theory, namely assuming their validity far beyond their 
empirically explored domain of validity, he predicted the 
existence of antimatter.

The two embryonal QFT's of Dirac were combined in the fifties by 
Feynman and his colleagues, giving rise to quantum 
electrodynamics, the first nontrivial interacting QFT. A 
remarkable picture of the world was born: quantum fields over 
Minkowski space.  Equivalently, \`a la Feynman: the world as a 
quantum superposition of histories of real and virtual 
interacting particles.  QFT had ups and downs, then triumphed with 
the standard model: a consistent QFT for all interactions (except 
gravity), which, in principle, can be used to predict anything we 
can measure (except gravitational phenomena), and which, in the 
last fifteen years has received nothing but empirical 
verifications.  

\subsection{Third step. The stage becomes an actor}

Descartes, in {\em Le Monde}, gave a fully relational definition 
of localization (space) and motion (on the 
relational/substantivalist issue, see Earman and Norton 1987, 
Barbour 1989, Earman 1989, Rovelli 1991a, Belot 1998).  According 
to Descartes, there is no ``empty space''.  There are only 
objects, and it makes sense to say that an object A is contiguous 
to an object B. The ``location'' of an object A is the set of the 
objects to which A is contiguous.  ``Motion'' is change in 
location.  That is, when we say that A moves we mean that A goes 
from the contiguity of an object B to the contiguity of an object 
C\footnote{``We can say that movement is the transference of one 
part of matter or of one body, from the vicinity of those bodies 
immediately contiguous to it, and considered at rest, into the 
vicinity of some others'', (Descartes, {\em Principia 
Philosophiae}, Sec II-25, pg 51).}.  A consequence of this 
relationalism is that there is no meaning in saying ``A moves'', 
except if we specify with respect to which other objects (B, 
C,\ldots) it is moving.  Thus, there is no ``absolute'' motion.  
This is the same definition of space, location, and motion, that 
we find in Aristotle.  \footnote{Aristotle insists on this point, 
using the example of the river that moves with respect to the 
ground, in which there is a boat that moves with respect to the 
water, on which there is a man that walks with respect to the 
boat \ldots.  Aristotle's relationalism is tempered by the fact 
that there is, after all, a preferred set of objects that we can 
use as universal reference: the Earth at the center of the 
universe, the celestial spheres, the fixed stars.  Thus, we can 
say, if we desire so, that something is moving ``in absolute 
terms'', if it moves with respect to the Earth.  Of course, there 
are {\em two\/} preferred frames in ancient cosmology: the one of 
the Earth and the one of the fixed stars; the two rotates with 
respect to each other.  It is interesting to notice that the 
thinkers of the middle ages did not miss this point, and 
discussed whether we can say that the stars rotate around the 
Earth, rather than being the Earth that rotates under the {\em 
fixed\/} stars.  Buridan concluded that, on ground of reason, in 
no way one view is more defensible than the other.  For 
Descartes, who writes, of course, after the great Copernican 
divide, the Earth is not anymore the center of the Universe and 
cannot offer a naturally preferred definition of stillness.  
According to malignants, Descartes, fearing the Church and scared 
by what happened to Galileo's stubborn defense of the idea that 
``the Earth moves'', resorted to relationalism, in {\em Le 
Monde}, precisely to be able to hold Copernicanism without having 
to commit himself to the absolute motion of the Earth!}

Relationalism, namely the idea that motion can be defined only in 
relation to other objects, should not be confused with Galilean 
relativity.  Galilean relativity is the statement that 
``rectilinear uniform motion'' is a priori indistinguishable from 
stasis.  Namely that velocity (but just velocity!), is relative 
to other bodies.  Relationalism holds that {\em any\/} motion 
(however zigzagging) is a priori indistinguishable from stasis.  
The very formulation of Galilean relativity requires a 
nonrelational definition of motion (``rectilinear and uniform'' 
with respect to what?).

Newton took a fully different course.  He devotes much energy to 
criticise Descartes' relationalism, and to introduce a different 
view.  According to him, {\em space\/} exists.  It exists even if 
there are no bodies in it.  Location of an object is the part of 
space that the object occupies.  Motion is change of 
location.\footnote{
``So, it is necessary that the definition of places, and hence 
local motion, be referred to some motionless thing such as 
extension alone or {\em space\/}, in so far as space is seen 
truly distinct from moving bodies'', (Newton {\it De gravitatione et 
Aequipondio Fluidorum} 89-156). Compare with the 
quotation of Descartes in the footnote above.}  
Thus, we can say whether an object moves or not, irrespectively 
from surrounding objects.  Newton argues that the notion of 
absolute motion is necessary for constructing mechanics.  His 
famous discussion of the experiment of the rotating bucket in the 
{\em Principia\/} is one of the arguments to prove that motion is 
absolute.

This point has often raised confusion because one of the 
corollaries of Newtonian mechanics is that there is no detectable 
preferred referential frame.  Therefore the notion of {\em 
absolute velocity\/} is, actually, meaningless, in Newtonian 
mechanics.  The important point, however, is that in Newtonian 
mechanics velocity is relative, but any other feature of motion 
is not relative: it is absolute.  In particular, acceleration is 
absolute.  It is acceleration that Newton needs to construct his 
mechanics; it is acceleration that the bucket experiment is 
supposed to prove to be absolute, against Descartes.  In a sense, 
Newton overdid a bit, introducing the notion of absolute position 
and velocity (perhaps even just for explanatory purposes?).  Many 
people have later criticised Newton for his unnecessary use of 
absolute position.  But this is irrelevant for the present 
discussion.  The important point here is that Newtonian 
mechanics requires absolute acceleration, against Aristotle and 
against Descartes.  Precisely the same does special relativistic 
mechanics.

Similarly, Newton introduce absolute time.  Newtonian space and 
time or, in modern terms, spacetime, are like a {\em stage\/} 
over which the action of physics takes place, the various 
dynamical entities being the actors.  

The key feature of this stage, Newtonian spacetime, is its 
metrical structure.  Curves have length, surfaces have area, 
regions of spacetime have volume.  Spacetime points are at fixed 
{\em distance\/} the one from the other.  Revealing, or 
measuring, this distance, is very simple.  It is sufficient to 
take a rod and put it between two points.  Any two points which 
are one rod apart are at the same distance.  Using modern 
terminology, physical space is a linear three-dimensional (3d) 
space, with a preferred metric.  On this space there exist 
preferred coordinates $x^{i},\ i=1,2,3$, in terms of which the 
metric is just $\delta_{ij}$.  Time is described by a single 
variable $t$.  The metric $\delta_{ij}$ determines lengths, areas 
and volumes and defines what we mean by straight lines in space.  
If a particle deviates with respect to this straight line, it is, 
according to Newton, accelerating.  It is not accelerating with 
respect to this or that dynamical object: it is accelerating in 
absolute terms.

Special relativity changes this picture only marginally, loosing 
up the strict distinction between the ``space'' and the ``time'' 
components of spacetime.  In Newtonian spacetime, space is given 
by fixed 3d planes.  In special relativistic spacetime, which 3d 
plane you call space depends on your state of motion.  Spacetime 
is now a 4d manifold $M$ with a flat Lorentzian metric 
$\eta_{\mu\nu}$.  Again, there are preferred coordinates 
$x^{\mu},\ \mu=0,1,2,3$, in terms of which $\eta _{\mu\nu} = 
diag[1,-1,-1,-1]$.  This tensor, $\eta_{\mu\nu}$, enters all 
physical equations, representing the determinant influence of the 
stage and of its metrical properties on the motion of anything.  
Absolute acceleration is deviation of the world line of a 
particle from the straight lines defined by $\eta_{\mu\nu}$.  The 
only essential novelty with special relativity is that the 
``dynamical objects'', or ``bodies'' moving over spacetime now 
include the fields as well.  Example: a violent burst of 
electromagnetic waves coming from a distant supernova has {\em 
traveled across space\/} and has reached our instruments.  For 
the rest, the Newtonian construct of a fixed background stage 
over which physics happen is not altered by special relativity.

The profound change comes with general relativity (GR).  The 
central discovery of GR, can be enunciated in three points.  One 
of these is conceptually simple, the other two are tremendous.  
First, the gravitational force is mediated by a field, very much 
like the electromagnetic field: the gravitational field.  Second, 
Newton's {\em spacetime}, the background stage that Newton 
introduced introduced, against most of the earlier European 
tradition, {\em and the gravitational field, are the same thing}.  
Third, the dynamics of the gravitational field, of the other 
fields such as the electromagnetic field, and any other dynamical 
object, is fully relational, in the Aristotelian-Cartesian sense.  
Let me illustrate these three points.

First, the gravitational field is represented by a field on 
spacetime, $g_{\mu\nu}(x)$, just like the electromagnetic field 
$A_{\mu}(x)$.  They are both very concrete entities: a strong 
electromagnetic wave can hit you and knock you down; and so can a 
strong gravitational wave.  The gravitational field has 
independent degrees of freedom, and is governed by dynamical 
equations, the Einstein equations.

Second, the spacetime metric $\eta_{\mu\nu}$ disappears from all 
equations of physics (recall it was ubiquitous).  At its place 
--we are instructed by GR-- we must insert the gravitational 
field $g_{\mu\nu}(x)$.  This is a spectacular step: Newton's 
background spacetime was nothing but the gravitational field!  
The stage is promoted to be one of the actors.  Thus, in all 
physical equations one now sees the direct influence of the 
gravitational field.  How can the gravitational field determine 
the metrical properties of things, which are revealed, say, by 
rods and clocks?  Simply, the inter-atomic separation of the rods' 
atoms, and the frequency of the clock's pendulum are determined 
by explicit couplings of the rod's and clock's variables with the 
gravitational field $g_{\mu\nu}(x)$, which enters the equations 
of motion of these variables.  Thus, any measurement of length, 
area or volume is, in reality, a measurement of features of the 
gravitational field. 

But what is really formidable in GR, the truly momentous 
novelty, is the third point: the Einstein equations, as well as 
{\em all other equations of physics} appropriately modified 
according to GR instructions, are fully relational in the 
Aristotelian-Cartesian sense.  This point is independent from the 
previous one.  Let me give first a conceptual, then a technical 
account of it.

The point is that the only physically meaningful definition of 
location that makes physical sense within GR is relational.  GR 
describes the world as a set of interacting fields and, possibly, 
other objects.  One of these interacting fields is 
$g_{\mu\nu}(x)$.  Motion can be defined only as positioning and 
displacements of these dynamical objects relative to each other 
(for more details on this, see Rovelli 1991a and especially 
1997a).

To describe the motion of a dynamical object, Newton had to 
assume that acceleration is absolute, namely it is not relative 
to this or that other dynamical object.  Rather, it is relative 
to a background space.  Faraday Maxwell and Einstein extended the 
notion of ``dynamical object'': the stuff of the world is fields, 
not just bodies.  Finally, GR tells us that the background space 
is itself one of these fields.  Thus, the circle is closed, and 
we are back to relationalism: Newton's motion with respect to 
space is indeed motion with respect to a dynamical object: the 
gravitational field.

All this is coded in the active diffeomorphism invariance (diff 
invariance) of GR.\footnote{Active diff invariance should not be 
confused with passive diff invariance, or invariance under change 
of coordinates.  GR can be formulated in a coordinate free 
manner, where there are no coordinates, and no changes of 
coordinates.  In this formulation, there field equations are {\em 
still\/} invariant under active diffs.  Passive diff invariance 
is a property of a formulation of a dynamical theory, while 
active diff invariance is a property of the dynamical theory 
itself.  A field theory is formulated in manner invariant under 
passive diffs (or change of coordinates), if we can change the 
coordinates of the manifold, re-express all the geometric 
quantities (dynamical {\em and non-dynamical\/}) in the new 
coordinates, and the form of the equations of motion does not 
change.  A theory is invariant under active diffs, when a smooth 
displacement of the dynamical fields ({\em the dynamical fields 
alone\/}) over the manifold, sends solutions of the equations of 
motion into solutions of the equations of motion.  Distinguishing 
a truly dynamical field, namely a field with independent degrees 
of freedom, from a nondynamical filed disguised as dynamical 
(such as a metric field $g$ with the equations of motion 
Riemann[g]=0) might require a detailed analysis (for instance, 
hamiltonian) of the theory.} Because active diff invariance is a 
gauge, the physical content of GR is expressed only by those 
quantities, derived from the basic dynamical variables, which are 
fully independent from the points of the manifold.  

In introducing the background stage, Newton introduced two 
structures: a spacetime manifold, and its non-dynamical metric 
structure.  GR gets rid of the non-dynamical metric, by replacing 
it with the gravitational filed.  More importantly, it gets rid 
of the manifold, by means of active diff invariance.  In GR, the 
objects of which the world is made do not live over a stage and 
do not live on spacetime: they live, so to say, over each 
other's shoulders.

Of course, nothing prevents us, if we wish to do so, from 
singling out the gravitational field as ``the more equal among 
equals'', and declaring that location is absolute in GR, because 
it can be defined with respect to it.  But this can be done 
within any relationalism: we can always single out a set of 
objects, and declare them as not-moving by 
definition\footnote{Notice that Newton, in the passage quoted in 
the footnote above argues that motion must be defined with respect 
to motionless space ``in so far as space is seen truly distinct 
from moving bodies''. That is: motion should be defined with 
respect to something that has no dynamics.}.  The problem with 
this attitude is that it fully misses the great Einsteinian 
insight: that Newtonian spacetime is just one field among the 
others.  More seriously, this attitude sends us into a nightmare 
when we have to deal with the motion of the gravitational field 
itself (which certainly ``moves'': we are spending millions for 
constructing gravity wave detectors to detect its tiny 
vibrations).  There is no absolute referent of motion in GR: the 
dynamical fields ``move'' with respect to each other.

Notice that the third step was not easy for Einstein, and came 
later than the previous two.  Having well understood the first 
two, but still missing the third, Einstein actively searched for 
non-generally covariant equations of motion for the gravitational 
field between 1912 and 1915.  With his famous ``hole argument'' 
he had convinced himself that generally covariant equations of 
motion (and therefore, in this context, active diffeomorphism 
invariance) would imply a truly dramatic revolution with respect 
to the Newtonian notions of space and time (on the hole argument, 
see Earman and Norton 1987, Rovelli 1991a, Belot 1998).  In 1912 
he was not able to take this profoundly revolutionary step 
(Norton 1984, Stachel 1989).  In 1915 he took this step, and 
found what Landau calls ``the most beautiful of the physical 
theories''.

\subsection{Bringing the three steps together}

At the light of the three steps illustrated above, the task of 
quantum gravity is clear and well defined.  He have learned from 
GR that spacetime is a dynamical field among the others, obeying 
dynamical equations, and having independent degrees of freedom.  A 
gravitational wave is extremely similar to an electromagnetic 
wave.  We have learned from QM that every dynamical object has 
quantum properties, which can be captured by appropriately 
formulating its dynamical theory within the general scheme of QM. 

{\em Therefore\/}, spacetime itself must exhibit quantum 
properties.  Its  properties, including the  metrical 
properties it defines, must be represented in quantum mechanical 
terms.  Notice that the strength of this ``therefore'' derives 
from the confidence we have in the two theories, QM and GR.

Now, there is nothing in the basics of QM which contradicts the 
physical ideas of GR. Similarly, there is nothing in the basis of 
GR that contradicts the physical ideas of QM. Therefore, there is 
no a priori impediment in searching for a quantum theory of the 
gravitational fields, that is, a quantum theory of spacetime.  
The problem is (with some qualification) rather well posed: is 
there a quantum theory (say, in one formulation, a Hilbert space 
$H$, and a set of self-adjoint operators) whose classical limit 
is GR? 

On the other hand, all previous applications of QM to {\em 
field\/} theory, namely conventional QFT's, rely heavily on the 
existence of the ``stage'', the fixed, non-dynamical, background 
metric structure.  The Minkowski metric $\eta_{\mu\nu}$is 
essentially for the construction of a conventional QFT (in enters 
everywhere; for instance, in the canonical commutation relations, 
in the propagator, in the Gaussian measure \ldots).  We certainly 
cannot simply replace $\eta_{\mu\nu}$ with a quantum field, 
because all equations become nonsense.  

Therefore, to search for a quantum theory of gravity, we have two 
possible directions.  One possibility is to ``disvalue'' the GR 
conceptual revolution, reintroduce a background spacetime with a 
non-dynamical metric $\eta_{\mu\nu}$, expand the gravitational 
field $g_{\mu\nu}$ as $g_{\mu\nu} = \eta_{\mu\nu} + 
fluctuations$, quantize only the fluctuations, and hope to 
recover the full of GR somewhere down the road.  This is the road 
followed for instance by perturbative string theory. 

The second direction is to be faithful to what we have learned 
about the world so far.  Namely to the QM and the GR insights.  
We must then search a QFT that, genuinely, does not 
require a background space to be defined.  But the last three 
decades whave been characterized by the great success of 
conventional QFT, which neglects GR and is based on the existence 
of a background spacetime.  We live in the aftermath of this 
success.  It is not easy to get out from the mental habits and 
from the habits to the technical tools of conventional QFT. 
Still, this is necessary if we want to build a QFT which fully 
incorporates active diff invariance, and in which localization is 
fully relational.  In my opinion, this is the right way to go.

\section{Quantum spacetime}

\subsection{Space}

Spacetime, or the gravitational field, is a dynamical entity 
(GR).  All dynamical entities have quantum properties (QM).  
Therefore spacetime is a quantum object.  It must be described 
(picking one formulation of QM, but keeping in mind that others 
may be equivalent, or more effective) in terms of states $\Psi$ 
in a Hilbert space.  Localization is relational.  Therefore these 
states cannot represent quantum excitations localized in some 
space.  They must define space themselves.  They must be quantum 
excitations ``of'' space, not ``in'' space.  Physical quantities 
in GR, that capture the true degrees of freedom of the theory are 
invariant under active diff.  Therefore the 
self-adjoint operators that correspond to physical (predictable) 
observables in quantum gravity must be associated to 
diff invariant quantities.

Examples of diff-invariant geometric quantities are physical 
lengths, areas, volumes, or time intervals, of regions determined 
by dynamical physical objects.  These must be represented by 
operators.  Indeed, a measurement of length, area or volume is a 
measurement of features of the gravitational field.  If the 
gravitational field is a quantum field, then length, area and 
volume are quantum observables.  If the corresponding operator 
has discrete spectrum, they will be quantized, namely they can 
take certain discrete values only.  In this sense we should 
expect a discrete geometry.  This discreteness of the geometry, 
implied by the conjunction of GR and QM is very different from 
the naive idea that the world is made by discrete bits of 
something.  It is like the discreteness of the quanta of the 
excitations of an harmonic oscillator.  A generic state of 
spacetime will be a continuous quantum superposition of states 
whose geometry has discrete features, not a collection of 
elementary discrete objects.

A concrete attempt to construct such a theory, is loop quantum 
gravity.  I refer the reader to Rovelli (1997b) for an 
introduction to the theory, an overview of its structure and 
results, and full references.  Here, I present only a few remarks 
on the theory.  Loop quantum gravity is a rather straightforward 
application of quantum mechanics to hamiltonian general 
relativity.  It is a QFT in the sense that it is a quantum 
version of a field theory, or a quantum theory for an infinite 
number of degrees of freedom, but it is profoundly different from 
conventional, non-general-relativistic QFT theory.  In 
conventional QFT, states are quantum excitations of a field over 
Minkowski (or over a curved) spacetime.  In loop quantum gravity, 
the quantum states turn out to be represented by (suitable linear 
combinations of) spin networks (Rovelli and Smolin 1995a, Baez 
1996, Smolin 1997).  A spin network is an abstract graphs with 
links labeled by half-integers.  See Figure 1.

\begin{figure} 
\centerline{\mbox{\epsfig{file=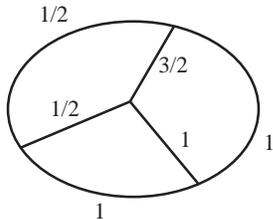}}} 
\caption{A simple spin network.}
\end{figure}

Intuitively, we can view each node of the graph as an elementary 
``quantum chunk of space''.  The links represent (transverse) 
surfaces separating the quanta of space.  The half-integers 
associated to the links determine the (quantized) area of these 
surfaces.  The spin network represent relational quantum states: 
they are not located in a space.  Localization must be defined in 
relation to them.  For instance, if we have, say, a matter 
quantum excitation, this will be located on the spin network; 
while the spin network itself is not located anywhere.

The operators corresponding to area and volume have been 
constructed in the theory, simply by starting from the classical 
expression for the area in terms of the metric, then replacing 
the metric with the gravitational field (this is the input of GR) 
and then replacing the gravitational field with the corresponding 
quantum field operator (this is the input of QM).  The 
construction of these operators requires appropriate generally 
covariant regularization techniques, but no renormalization: no 
infinities appear.  The spectrum of these operators has been 
computed and turns out to be discrete (Rovelli and Smolin 1995b, 
Ashtekar Lewandowski 1997a, 1997b).  Thus, loop quantum gravity 
provides a family of precise quantitative predictions: the 
quantized values of area and volume.  For instance, the (main 
sequence) of the spectrum of the area is
\begin{displaymath}
	A = 8\,\pi\,\hbar\,G \sum_{i=1,n}\sqrt{j_i(j_i+1)} 
\end{displaymath}
where $(j_{i})=(j_{1}\ldots j_{n})$ is any finite sequence of 
half integers.  This formula gives the area of a surface pinched 
by $n$ links of a spin network state.  The half integers 
$j_{1}\ldots j_{n}$ are ones associated with the $n$ links that 
pinch the surface.  This illustrates how the links of the spin 
network states can be viewed as transversal ``quanta of area''.  
The picture of macroscopic physical space that emerges is then 
that of a tangle of one-dimensional intersecting quantum 
excitation, called the weave (Ashtekar Rovelli and Smolin 1992).  
Continuous space is formed by the weave in the same manner in 
which the continuous 2d surface of a T-shirt is formed by weaved 
threads.

\subsection{Time}

The aspect of the GR's relationalism that concerns space was 
largely anticipated by the earlier European thinking.  Much less 
so (as far as I am aware) was the aspect of this relationalism 
that concerns time.  GR's treatment of time is surprising, 
difficult to fully appreciate, and hard to digest.  The time of 
our perceptions is very different from the time that theoretical 
physics finds in the world as soon as one exits the minuscule 
range of physical regimes we are accustomed to. We seem to 
have a very special difficulty in being open minded about this 
particular notion. 

Already special relativity teaches us something about time which 
many of us have difficulties to accept.  According to special 
relativity, there is absolute no meaning in saying ``right now on 
Andromeda''.  There is no physical meaning in the idea of ``the 
state of the world right now'', because which set of events we 
consider as ``now'' is perspectival.  The ``now'' on Andromeda 
for me might correspond to ``a century ago'' on Andromeda for 
you.  Thus, there is no single well defined universal time in 
which the history of the universe ``happens''.  The modification 
of the concept of time introduced by GR is much deeper.  Let me 
illustrate this modifications.

Consider a simple pendulum described by a variable $Q$.  In 
Newtonian mechanics, the motion of the pendulum is given by the 
evolution of $Q$ in time, namely by $Q(T)$, which is governed by 
the equation of motion, say $\ddot Q=-\omega Q$, which has (the 
two-parameter family of) solutions $Q(T)=A\sin(\omega T+\phi)$.  
The state of the pendulum at time $T$ can be characterized by its 
position and velocity.  From these two, we can 
compute $A$ and $\phi$ and therefore $Q(T)$ at any $T$.  From the 
physical point of view, we are really describing a situation in 
which there are {\em two\/} physical objects: a pendulum, whose 
position is $Q$, and a clock, indicating $T$.  If we want to take 
data, we have to repeatedly observe $Q$ and $T$.  Their {\em 
relation\/} will be given by the equation above.  The relation 
can be represented (for given $A$ and $\phi$) by a line in the 
$(Q,T)$ plane.

In Newtonian terms, time flows in its absolute way, the clock is 
just a devise to keep track of it, and the dynamical system is 
formed by the pendulum alone.  But we can view the same physical 
situation from a different perspective.  We can say that we have 
a physical system formed by the clock and the pendulum together 
and view the dynamical system as expressing the relative motion 
of one with respect to the other.  This is precisely the 
perspective of GR: to express the relative motion of the 
variables, with respect to each other, in a ``democratic'' 
fashion.

To do that, we can introduce an ``arbitrary parameter time'' 
$\tau$ as a coordinate on the line in the $(Q,T)$ plane.  (But 
keep in mind that the physically relevant information is in the 
line, not in its coordinatization!).  Then the line is 
represented by two functions, $Q(\tau)$ and $T(\tau)$, but a 
reparametrization of $\tau$ in the two functions is a gauge, 
namely it does not modify the physics described.  Indeed, $\tau$ 
does not correspond to anything observable, and the equations of 
motion satisfied by $Q(\tau)$ and $T(\tau)$ (easy to write, but I 
will not write them down here) will be invariant under arbitrary 
reparametrizations of $\tau$.  Only $\tau$-independent quantities 
have physical meaning.

This is precisely what happens in GR, where the ``arbitrary 
parameters'', analogous to the $\tau$ of the example, are the 
coordinates $x^\mu$.  Namely, the spatial coordinate $\vec x$ and 
the temporal coordinate $t$.  These have no physical meaning 
whatsoever in GR: the connection between the theory and the 
measurable physical quantities that the theory predict is only 
via quantities independent from $\vec x$ and $t$.  Thus, $\vec x$ 
and $t$ in GR have a very different physical meaning than their 
homonymous in non-general-relativistic physics.  The later 
correspond to readings on rods and clocks.  The formed, 
correspond to nothing at all.  Recall that Einstein described his 
great intellectual struggle to find GR as ``understanding the 
meaning of the coordinates''.

In the example, the invariance of the equations of motion for 
$Q(\tau)$ and $T(\tau)$ under reparametrization of $\tau$, 
implies that if we develop the Hamiltonian formalism in $\tau$ we 
obtain a constrained system with a (weakly) vanishing 
hamiltonian.  This is because the hamiltonian generates 
evolutions in $\tau$, evolution in $\tau$ is a gauge, and the 
generators of gauge transformations are constraints.  In 
canonical GR we have precisely the same situation: the hamiltonian 
vanishes, the constraints generate evolution in $t$, which is 
unobservable -- it is gauge.  GR does not describe evolution in time: 
it describes the relative evolution of many variables with 
respect to each other.  All these variables are democratically 
equal: there isn't a preferred one that ``is the true time''.  
This is the temporal aspect of GR's relationalism.

A large part of the machinery of theoretical physics relies on 
the notion of time (on the different meanings of time in 
different physical theories, see Rovelli 1995).  A theory of 
quantum gravity should do without.  Fortunately, many essential 
tools that are usually introduced using the notion of time can 
equally well be defined without mentioning time at all.  This, by 
the way, shows that time plays a much weaker role in the 
structure of theoretical physics than what is mostly assumed.  
Two crucial examples are ``phase space'' and ``state''. 

The phase space is usually introduced in textbooks as the space 
of the states of the systems ``at a given time''.  In a general 
relativistic context, this definition is useless.  However, it 
is known since Lagrange that there is an alternative, equivalent,  
definition of phase space as the space of the solutions of the 
equations of motion.  This definition does not require that we 
know what we mean by time.  Thus, in the example above the phase 
space can be coordinatized by $A$ and $\phi$, which coordinatize 
the space of the solutions of the equations of motion.  

A time independent notion of ``state'' is then provided by a 
point of this phase space, namely by a particular solution of the 
equations of motion.  For instance, for an oscillator a 
``state'', in this atemporal sense, is characterized by an 
amplitude $A$ and a phase $\phi$.  Notice that given the 
(time-independent) state ($A$ and $\phi$), we can compute any 
observable: in particular, the value $Q_{T}$ of $Q$ at any 
desired $T$.  Notice also that $Q_{T}$ is independent from 
$\tau$.  This point often raises confusion: one may think that if 
we restrict to $\tau$-independent quantities then we cannot 
describe evolution.  This is wrong: the true evolution is the 
relation between $Q$ and $T$, which is $\tau$-independent.  This 
relation is expressed in particular by the value (let us denote 
it $Q_{T}$) of $Q$ at a given $T$.  $Q_{T}$ is given, obviously,  
by 
\begin{displaymath}  
       Q_{T}(A,\phi)=A \sin(\omega T+\phi).
\end{displaymath}
This can be seen as a one-parameter (the parameter is $T$) family 
of observables on the gauge invariant phase space coordinatized 
by $A$ and $\phi$.  Notice that this is a perfectly 
$\tau$-independent expression.  In fact, an explicit computation 
shows that the Poisson bracket between $Q_{T}$ and the 
hamiltonian constraint that generates evolution in $\tau$ 
vanishes.

This time independent notion of states is well known in its 
quantum mechanical version: it is the Heisenberg state (as 
opposed to Schr\"odinger state).  Similarly, the operator 
corresponding to the observable $Q_{T}$ is the Heisenberg 
operator that gives the value of $Q$ at $T$.  The Heisenberg and 
Schr\"odinger pictures are equivalent if there is a normal time 
evolution in the theory.  In the absence of a normal notion of 
time evolution, the Heisenberg picture remains viable, the 
Shr\"odinger picture becomes meaningless.\footnote{In the first 
edition of his celebrated book on quantum mechanics, Dirac used 
Heisenberg states (he calls them relativistic).  In later 
editions, he switched to Shr\"odinger states, explaining in a 
preface that it was easier to calculate with these, but it was 
nevertheless a pity to give up the Heisenberg states, which are 
more fundamental.  In what was perhaps his last public seminar, 
in Sicily, Dirac used just a single transparency, with just one 
sentence: ``The Heisenberg picture is the right one''.} In 
quantum gravity, only the Heisenberg picture makes sense (Rovelli 
1991c, 1991d).

In classical GR, a point in the physical phase space, or a state, 
is a solution of Einstein equations, up to active 
diffeomorphisms.  A state represents a ``history'' of spacetime.  
The quantity that can be univocally predicted are the ones that 
are independents from the coordinates, namely that are invariant 
under diffeomorphisms.  These quantities have vanishing Poisson 
brackets with all the constraints.  Given a state, the value of 
each of these quantities is determined.  In quantum gravity, a 
quantum state represents a ``history'' of quantum spacetime.  The 
observables are represented by operators that commute with {\em 
all\/} the quantum constraints.  If we know the quantum state of 
spacetime, we can then compute the expectation value of any 
diffeomorphism invariant quantity, by taking the mean value of 
the corresponding operator.  The observable quantities in quantum 
gravity are precisely the same as in classical GR.

Some of these quantities may express the value of certain 
variables ``when and where'' certain other quantities have 
certain given values.  They are the analog of the 
reparametrization invariant observable $Q_{T}$ in the example 
above.  These quantities describe evolution in a way which is 
fully invariant under the parameter time, unphysical gauge 
evolution (Rovelli 1991d, 1991e).  The corresponding quantum 
operators are Heisenberg operators.  There is no Schr\"odinger 
picture, because there is no unitary time evolution.  There is no 
need to expect or to search for unitary time evolution in quantum 
gravity, because there is no time in which we should have unitary 
evolution.  A prejudice hard to die wants that unitary evolution 
is required for the consistency of the probabilistic 
interpretation.  This idea is wrong.

What I have described is the general form that one may expect a 
quantum theory of GR to have.  I have used the Hilbert space 
version of QM; but this structure can be translated in other 
formulations of QM. Of course, physics works then with dirty hands: 
gauge dependent quantities, approximations, expansions, 
unphysical structures, and so on.  A fully satisfactory 
construction of the above does not yet exist.  A concrete attempt 
to construct the physical states and the physical observables in 
loop quantum gravity is given by the spin foam models approach, 
which is the formulation one obtains by starting from loop 
quantum gravity and constructing a Feynman sum over histories  
(Reisenberger Rovelli 1997, Baez 1998, Barret and Crane 1998).  
See (Baez 1999) in this volume for more details on ideas 
underlying these developments.

In quantum gravity, I see no reason to expect a fundamental 
notion of time to play any role.  But the {\em nostalgia for 
time\/} is hard to resist.  For technical as well as for 
emotional reasons.  Many approaches to quantum gravity go out of 
their way to reinsert in the theory what GR is teaching us we 
should abandon: a preferred time.  The time ``along which'' 
things happen is a notion which makes sense only for describing a 
limited regime of reality.  This notion is meaningless already in 
the (gauge invariant) general relativistic classical dynamics of 
the gravitational field.  At the fundamental level, we should, 
simply, forget time.

\subsection{Glimpses}

I close this section by briefly mentioning two more speculative ideas.  
One regards the emergence of time, the second the connection 
between the relationalism in GR and the relationalism in QM.

(i) In the previous section, I have argued that we should search 
for a quantum theory of gravity in which there is no independent 
time variable ``along'' which dynamics ``happens''.  A problem 
left open by this position is to understand the emergence of time 
in our world, with its features, which are familiar to us.  An 
idea discussed in (Rovelli 1993a 1993b, Connes and Rovelli 1994) 
is that the notion of time isn't dynamical but rather 
thermodynamical.  We can never give a complete account of the 
state of a system in a field theory (we cannot access the 
infinite amount of data needed to completely characterize a 
state).  Therefore we have at best a statistical description of 
the state.  Given a statistical state of a generally covariant 
system, a notion of a flow (more precisely a one-parameter group 
of automorphisms of the algebra of the observables) follows 
immediately.  In the quantum context, this corresponds to the 
Tomita flow of the state.  The relation between this flow and the 
state is the relation between the time flow generated by the 
hamiltonian and a Gibbs state: the two essentially determine each 
other.  In the absence of a preferred time, however, any 
statistical state selects its own notion of statistical time.  
This statistical time has a striking number of properties that 
allow us to identify it with the time of non-general relativistic 
physics.  In particular, a Schr\"odinger equation with respect to 
this statistical time holds, in an appropriate sense.  In 
addition, the time flows generated by different states are 
equivalent up to inner automorphisms of the observable algebra 
and therefore define a common ``outer'' flow: a one paramater 
group of outer automorphisms.  This determines a state 
independent notion of time flow, which shows that a general 
covariant QFT has an intrinsic ``dynamics'', even in the absence 
of a hamiltonian and of a time variable.  The suggestion is 
therefore that the temporal aspects of our world have statistical 
and thermodynamical origin, rather than dynamical.  ``Time'' is 
ignorance: a reflex of our incomplete knoweldge of the state of 
the world.

(ii) What is QM really telling us about our world?  In (Rovelli 
1996, 1998), I have argued that what QM is telling us is that the 
contingent properties of any system --or: the state of any 
system-- must be seen as relative to a second physical system, 
the ``observing system''.  That is, quantum state and values that 
an observables take are relational notions, in the same sense in 
which velocity is relational in classical mechanics (it is a 
relation between two systems, not a properties of a single 
system).  I find the consonance between this relationalism in QM 
and the relationalism in GR quite striking.  It is tempting to 
speculate that they are related.  Any quantum interaction (or 
quantum measurement) involving a system $A$ and a system $B$ 
requires $A$ and $B$ to be spatiotemporally contiguous.  
Viceversa, spatiotemporal contiguity, which is the grounding of 
the notions of space and time (derived and dynamical, not 
primary, in GR) can only be verified quantum mechanically (just 
because any interaction is quantum mechanical in nature).  Thus, 
the net of the quantum mechanical elementary interactions and the 
spacetime fabric are actually the same thing.  Can we build a 
consistent picture in which we take this fact into account?  To 
do that, we must identify two notions: the notion of a 
spatiotemporal (or spatial?)  region, and the notion of quantum 
system.  For intriguing ideas in this direction, see (Crane 1991) 
and, in this volume, (Baez 1999).

\section{Considerations on method and content}

\subsection{Method}

Part of the recent reflection about science has emphasized the 
``non cumulative'' aspect in the development of scientific 
knowledge.  According to this view, the evolution of scientific 
theories is marked by large or small breaking points, in which, 
to put it very crudely, the empirical facts are just reorganized 
within new theories.  These would be to some extent 
``incommensurable'' with respect to their antecedent.  These 
ideas have influenced physicists.

The reader has remarked that the discussion of quantum gravity I 
have given above assumes a different reading of the evolution 
of scientific knowledge.  I have based the above discussion on 
quantum gravity on the idea that the central physical ideas of QM 
and GR represent our best guide for accessing the extreme and 
unexplored territories of the quantum-gravitational regime.  In 
my opinion, the emphasis on the incommensurability between 
theories has probably clarified an important aspect of science, 
but risks to obscure something of the internal logic according to 
which, historically, physics finds knowledge.  There is a subtle, 
but definite, cumulative aspect in the progress of physics, which 
goes far beyond the growth of validity and precision of the 
empirical content of the theories.  In moving from a theory to 
the theory that supersedes it, we do not save just the verified 
empirical content of the old theory, but more.  This ``more'' is 
a central concern for good physics.  It is the source, I think, 
of the spectacular and undeniable predicting power of theoretical 
physics.

Let me illustrate the point I am trying to make with a historical 
case.  There was a problem between Maxwell equations and Galilei 
transformations.  There were two obvious way out.  To disvalue 
Maxwell theory, degrading it to a phenomenological theory of some 
yet-to-be-discovered ether's dynamics.  Or to disvalue Galilean 
invariance, accepting the idea that inertial systems are not 
equivalent in electromagnetic phenomena.  Both ways were pursued 
at the end of the century.  Both are sound applications of the 
idea that a scientific revolution may very well change in depth 
what old theories teach us about the world.  Which of the two 
ways did Einstein take?

None of them.  For Einstein, Maxwell theory was a source of great 
awe.  Einstein rhapsodizes about his admiration for Maxwell 
theory.  For him, Maxwell had opened a new window over the world.  
Given the astonishing success of Maxwell theory, empirical 
(electromagnetic waves), technological (radio) as well as 
conceptual (understanding what is light), Einstein admiration is 
comprehensible.  But Einstein had a tremendous respect for 
Galileo's insight as well.  Young Einstein was amazed by a book 
with Huygens' derivation of collision theory virtually out of 
Galilean invariance alone.  Einstein understood that Galileo's 
great intuition --that the notion of velocity is only relative-- 
{\it could not be wrong}.  I am convinced that in this faith of 
Einstein in the core of the great Galilean discovery there is 
very much to learn, for the philosophers of science, as well as 
for the contemporary theoretical physicists.  So, Einstein {\em 
believed the two theories, Maxwell and Galileo}.  He assumed that 
they would hold far beyond the regime in which they had been 
tested.  He assumed that Galileo had grasped something about the 
physical world, which was, simply, {\em correct}.  And so had 
Maxwell.  Of course, details had to be adjusted.  The core of 
Galileo's insight was that all inertial systems are equivalent 
and that velocity is relative, not the details of the galilean 
transformations.  Einstein knew the Lorentz transformations 
(found, of course, by Lorentz, not by Einstein), and was able to 
see that they do not contradict Galileo's insight.  If there was 
contradiction in putting the two together, the problem was ours: 
we were surreptitiously sneaking some incorrect assumption into 
our deductions.  He found the incorrect assumption, which, of 
course, was that simultaneity could be well defined.  It was 
Einstein's faith in the {\em essential physical correctness\/} of 
the old theories that guided him to his spectacular discovery.

There are innumerable similar examples in the history of physics, 
that equally well could illustrate this point.  Einstein found GR 
``out of pure thought'', having Newton theory on the one hand and 
special relativity --the understanding that any interaction is 
mediated by a field-- on the other; Dirac found quantum field 
theory from Maxwell equations and quantum mechanics; Newton 
combined Galileo's insight that acceleration governs dynamics 
with Kepler's insight that the source of the force that governs 
the motion of the planets is the sun \ldots\ The list could be 
long.  In all these cases, confidence in the insight that came 
with some theory, or ``taking a theory seriously'', lead to major 
advances that largely extended the original theory itself.  Of 
course, far from me suggesting that there is anything simple, or 
automatic, in figuring out where the true insights are and in 
finding the way of making them work together.  But what I am 
saying is that figuring out where the true insights are and 
finding the way of making them work together is the work of 
fundamental physics.  This work is grounded on the {\em 
confidence\/} in the old theories, not on random search of new 
ones.

One of the central concerns of modern philosophy of science is 
to face the apparent paradox that scientific theories change, but 
are nevertheless credible.  Modern philosophy of science is to 
some extent an after-shock reaction to the fall of Newtonian 
mechanics.  A tormented recognition that an extremely successful 
scientific theory can nevertheless be untrue.  But it is a 
narrow-minded notion of truth the one which is questioned by the 
event of a successful physical theory being superseded by a more 
successful one.

A physical theory, in my view, is a conceptual structure that we 
use in order to organize, read and understand the world, and 
make  prediction about it.  A successful physical theory is a 
theory that does so effectively and consistently.  At the light 
of our experience, there is no reason not to expect that a more 
effective conceptual structure might always exist.  Therefore an 
effective theory may always show its limits and be replaced by a 
better one.  On the other hand, however, a novel 
conceptualization cannot but rely on what the previous one has 
already achieved.

When we move to a new city, we are at first confused about its 
geography.  Then we find a few reference points, and we make a 
rough mental map of the city in terms of these points.  Perhaps 
we see that there is part of the city on the hills and part on 
the plane.  As time goes on, the map gets better.  But there are 
moments, in which we suddenly realize that we had it wrong.  
Perhaps there were indeed two areas with hills, and we were 
previously confusing the two.  Or we had mistaken a big red 
building for the City Hall, when it was only a residential 
construction.  So we adjourn the mental map.  Sometime later, we 
have learned names and features of neighbors and streets; and the 
hills, as references, fade away.  The neighbors structure of 
knowledge is more effective that the hill/plane one \ldots 
The structure changes, but the knowledge increases.  And the big 
red building, now we know it, is not the City Hall, and we know 
it forever.

There are discoveries that are forever.  That the Earth is not 
the center of the universe, that simultaneity is relative.  That 
we do not get rain by dancing.  These are steps humanity takes, 
and does not take back.  Some of these discoveries amount simply to 
cleaning our thinking from wrong, encrusted, or provisional 
credences.  But also discovering classical mechanics, or 
discovering electromagnetism, or quantum mechanics, are 
discoveries forever.  Not because the details of these theories 
cannot change, but because we have discovered that a large 
portion of the world admits to be understood in certain terms, 
and this is a {\em fact\/} that we will have to keep facing 
forever.

One of the thesis of this essay, is that general relativity is 
the expression of one of these insights, which will stay with us 
``forever''.  The insight is that the physical world does not 
have a stage, that localization and motion are relational only, 
that diff-invariance (or something physically analogous) is 
required for any fundamental description of our world.

How can a theory be effective even outside the domain for which 
it was found?  How could Maxwell predict radio waves, Dirac 
predict antimatter and GR predict black holes?  How can 
theoretical thinking be so magically powerful?  Of course, we may 
think that these successes are chance, and historically deformed 
perspective.  There are hundred of theories proposed, most of 
them die, the ones that survive are the ones remembered.  There 
is alway somebody who wins the lottery, but this is not a sign 
that humans can magically predict the outcome of the lottery.  My 
opinion is that such an interpretation of the development of 
science is unjust, and, worse, misleading.  It may explain 
something, but there is more in science.  There are tens of 
thousand of persons playing the lottery, there were only two 
relativistic theories of gravity, in 1916, when Einstein 
predicted that the light would be defected by the sun precisely 
by an angle of 1.75''.  Familiarity with the history of physics, 
I feel confident to claim, rules out the lottery picture.

I think that the answer is simpler.  Somebody predicts that the 
sun will rise tomorrow, and the sun rises.  It is not a matter of 
chance (there aren't hundreds of people making random predictions 
on each sort of strange objects appearing at the horizon).  The 
prediction that tomorrow the sun will rise, is sound.  However, 
it is not granted either.  A neutron star could rush in, close to 
the speed of light, and sweep the sun away.  More 
philosophically, who grants me the right of induction?  Why 
should I be confident that the sun would rise, just because it 
has been rising so many time in the past?  I do not know the 
answer to {\em this\/} question.  But what I know is that the 
predictive power of a theory beyond its own domain is {\em 
precisely of the same sort.} Simply, we learn something about 
nature (whatever this mean).  And what we learn is effective in 
guiding us to predict nature's behavior.  Thus, the spectacular 
predictive power of theoretical physics is nothing less and 
nothing more than common induction.  And it is as comprehensible 
(or as incomprehensible) as my ability to predict that the sun 
will rise tomorrow.  Simply, nature around us happens to be full 
of regularities {\em that we understand\/}, whether or not we 
understand why regularities exist at all.  These regularities 
give us strong confidence -although not certainty- that the sun 
will rise tomorrow, as well as in the fact that the basic facts 
about the world found with QM and GR will be confirmed, not 
violated, in the quantum gravitational regimes that we have not 
empirically probed.

This view is not dominant nowadays in theoretical physics.  Other 
attitudes dominate.  The ``pragmatic'' scientist ignores 
conceptual questions and physical insights, and only cares about 
developing a theory.  This is an attitude, that has been 
successful in the sixties in getting to the standard model.  The 
``pessimistic'' scientist has little faith in the possibilities 
of theoretical physics, because he worries that all possibilities 
are open, and anything might happen between here and the Planck 
length.  The ``wild'' scientist observes that great scientists 
had the courage of breaking with old and respected ideas and 
assumptions, and explore new and strange hypothesis.  From this 
observation, the ``wild'' scientist concludes that to do great 
science one has to explore strange hypotheses, and {\em violate 
respected ideas\/}.  The wildest the hypothesis, the best. I 
think wilderness in physics is sterile.  The greatest 
revolutionaries in science were extremely, almost obsessively, 
conservative.  So was certainly the greatest revolutionary, 
Copernicus, and so was Planck.  Copernicus was pushed to the 
great jump from his pedantic labor on the minute technicalities 
of the Ptolemaic system (fixing the equant).  Kepler was forced 
to abandon the circles by his extremely technical work on the 
details of Mars orbit.  He was using ellipses as approximations 
to the epicycle-deferent system, when he begun to realize that 
the approximation was fitting the data better than the 
(supposedly) exact curve.  And extremely conservative were 
also Einstein and Dirac.  Their vertiginous steps ahead were not 
pulled out of the blue sky.  They did not come from violating 
respected ideas, but, on the contrary, from respect towards 
physical insights.  In physics, novelty has always emerged from 
new data and from a humble, devoted interrogation of the old 
theories. From turning these theories around and around, 
immerging into them, making them clash, merge, talk, until, 
through them, the missing gear could be seen.  In my opinion, 
precious research energies are today lost in these attitudes.  I 
worry that a philosophy of science that downplays the component 
of factual knowledge in physical theories might have part of the 
responsibility.

\subsection{On content and truth in physical theories}

If a physical theory is a conceptual structure that we use to 
organize, read and understand the world, then scientific thinking 
is not much different from common sense thinking.  In fact, it is 
only a better instance of the same activity: thinking about the 
world.  Science is the enterprise of continuously exploring the 
possible ways of thinking about the world, and constantly 
selecting the ones that work best.

If so, there cannot be any qualitative difference between the 
theoretical notions introduced in science and the terms in our 
everyday language.  A fundamental intuition of classical 
empiricism is that nothing grants us the ``reality'' of the 
referents of the notions we use to organize our perceptions.  
Some modern philosophy of science has emphasized the application 
of this intuition to the concepts introduced by science.  Thus, 
we are warned to doubt the ``reality'' of the theoretical objects 
(electrons, fields, black holes \ldots).  I find these warning 
incomprehensible.  Not because they are ill founded, but because 
they are not applied consistently.  The fathers of empiricism 
consistently applied this intuition to {\em any\/} physical 
object.  Who grants me the reality of a chair?  Why should a 
chair be more than a theoretical concept organizing certain 
regularities in my perceptions?  I will not venture here in 
disputing nor in agreeing with this doctrine.  What I find 
incomprehensible is the position of those who grant the solid 
status of reality to a chair, but not to an electron.  The 
arguments against the reality of the electron apply to the chair 
as well.  The arguments in favor of the reality of the chair 
apply to the electron as well.  A chair, as well as an electron, 
is a concept that we use to organize, read and understand the 
world.  They are equally real.  They are equally volatile and 
uncertain. 

Perhaps, this curious schizophrenic attitude of being antirealist 
with electrons and iron realist with chairs is the result of a 
complex historical evolution.  First there was the rebellion 
against ``metaphysics'', and, with it, the granting of confidence 
to science alone.  From this point of view, metaphysical 
questioning on the reality of chairs is sterile -- true knowledge 
is in science.  Thus, it is to scientific knowledge that we apply 
empiricist rigor.  But understanding science in empiricists' 
terms required making sense of the raw empirical data on which 
science is based.  With time, the idea of raw empirical data 
showed more and more its limits.  The common sense view of the 
world was reconsidered as a player in our picture of knowledge.  
This common sense view should give us a language and a ground 
from which to start -- the old anti-metaphysical prejudice still 
preventing us, however, from applying empiricist rigor to this 
common sense view of the world as well.  But if one is not 
interested in questioning the reality of chairs, for the very 
same reason why should one be interested in questioning the 
``reality of the electrons''?

Again, I think this point is important for science itself.  The 
factual content of a theory is our best tool.  The faith in this 
factual content does not prevent us from being ready to question 
the theory itself, if sufficiently compelled to do so by novel 
empirical evidence or by putting the theory in relation to other 
things {\em we know\/} about the world.  Scientific antirealism,  
in my opinion, is not only a short sighted application of a deep 
classical empiricist insight; it is also a negative influence 
over the development of science.  H.\ Stein (1999) has recently 
beautifully illustrated a case in which a great scientist, 
Poincar\'e, was blocked from getting to a major discovery 
(special relativity) by a philosophy that restrained him from 
``taking seriously'' his own findings.

Science teaches us that our naive view of the world is imprecise, 
inappropriate, biased.  It constructs better views of the world.  
Electrons, if anything at all, are ``more real'' that chairs, not 
``less real'', in the sense that they ground a more powerful way 
of conceptualizing the world.  On the other hand, the process of 
scientific discovery, and the experience of this century in 
particular, has made us painfully aware of the provisional 
character of {\em any\/} form of knowledge.  Our mental and 
mathematical pictures of the world are only mental and 
mathematical pictures.  This is true for abstract scientific 
theories as well as from the image we have of our dining room.  
Nevertheless, the pictures are powerful and effective and we 
can't do any better than that.

So, is there anything we can say with confidence about the ``real 
world''?  A large part of the recent reflection on science has 
taught us that row data do not exist, and that any information 
about the world is already deeply filtered and interpreted by the 
theory.  Further than that, we could even think, as in the dream 
of Berkeley, that there is no ``reality'' outside there.  The 
European reflection (and part of the American as well) has 
emphasized the fact that truth is always internal to the theory, 
that we can never exit language, we can never exit the circle of 
discourse within which we are speaking.  It might very well be 
so.  But, if the only notion of truth is internal to the theory, 
then {\em this internal truth\/} is what we mean by truth.  We 
cannot exit from our own conceptual scheme.  We cannot put 
ourself outside our discourse.  Outside our theory.  There may be 
no notion of truth outside our own discourse.  But it is 
precisely ``from within the language'' that we can assert the 
reality of the world.  And we certainly do so.  Indeed, it is 
more than that: it is structural to our language to be a language 
{\em about\/} the world, and to our thinking to be a thinking 
{\em of\/} the world.  Therefore, precisely because there is no 
notion of truth except the one in our own discourse, precisely 
for this reason, there is no sense in denying the reality of the 
world.  The world is real, solid, and understandable by science.  
The best we can say about the physical world, and about what is 
there in the world, is what good physics says about it.

At the same time, our perceiving, understanding, and 
conceptualizing the world is in continuous evolution, and science 
is the form of this evolution.  At every stage, the best we can 
say about the reality of the world is precisely what we are 
saying.  The fact we will understand it better later on does not 
make our present understanding less valuable, or less credible.  
A map is not false because there is a better map, even if the 
better one looks quite different.  Searching for a fixed point on 
which to rest our restlessness, is, in my opinion, naive, useless 
and counterproductive for the development of science.  It is only 
by believing our insights and, at the same time, questioning our 
mental habits, that we can go ahead.  This process of cautious 
faith and self-confident doubt is the core of scientific 
thinking.  Exploring the possible ways of thinking of the world, 
being ready to subvert, if required, our ancient prejudices, is 
among the greatest and the most beautiful of the human 
adventures.  Quantum gravity, in my view, in its effort to 
conceptualize quantum spacetime, and to modify in depth the 
notion of time, is a step of this adventure.

\vskip 2cm

\begin{itemize}

\item[]  
Ashtekar A, Rovelli C, Smolin L (1992). Weaving a classical 
metric with quantum threads, {\it Physical Review Letters}  {\bf 69}, 
237.  

\item[]
Ashtekar A , Lewandowski J (1997a). Quantum Theory of Gravity I:
Area Operators, {\it Class and Quantum Grav} {\bf 14}, A55--A81.

\item[]
Ashtekar A, Lewandowski J (1997b). Quantum Theory 
of Geometry II: Volume operators, gr-qc/9711031.

\item[]
Baez J (1997). Spin networks in nonperturbative quantum gravity, 
in {\it The interface of knots and physics}, ed L Kauffman 
(American Mathematical Society, Providence).

\item[]
Baez J (1998). Spin foam models, {\it Class Quantum Grav}, {\bf 15}, 
1827--1858.  

\item[]
Baez J (1999). In ``Physics Meets Philosophy at the Planck 
scale'', C Callender N Hugget eds, Cambridge University 
Press, to appear.

\item[]
Barbour J (1989). {\em Absolute or Relative Motion?} (Cambridge University 
Press, Cambridge).

\item[]
Barret J, Crane L (1998). Relativistic spin networks and quantum 
gravity, {\em Journ Math Phys} {\bf 39}, 3296--3302.

\item[]
Belot G (1998). Why general relativity does need an 
interpretation, {\it Philosophy of Science}, {\bf 63}, S80--S88.  

\item[]
Connes A and Rovelli C (1994). Von Neumann algebra automorphisms and 
time versus thermodynamics relation in general covariant quantum 
theories, {\it Classical and Quantum Gravity} {\bf 11}, 2899.

\item[]
Crane L (1991). 2d physics and 3d topology, {\em Comm Math 
Phys} {\bf 135}, 615--640.

\item[]
Descartes R (1983): {\em Principia Philosophiae}, 
Translated by VR Miller and RP Miller (Reidel, Dordrecht 
[1644]). 

\item[]
Earman J (1989). {\it World Enough and Spacetime: 
Absolute versus Relational Theories of Spacetime} (MIT Press, 
Cambridge).  

\item[]
Earman J, Norton J (1987). What Price Spacetime 
Substantivalism? The Hole Story, {\it British Journal for the 
Philosophy of Science}, {\bf 38}, 515--525.

\item[]
Isham C (1999). In ``Physics Meets Philosophy at the Planck 
scale'', C Callender N Hugget eds, Cambridge University 
Press, to appear.

\item[]
Newton I (1962): {\em De Gravitatione et Aequipondio 
Fluidorum}, translation in AR Hall and MB Hall eds {\it Unpublished 
papers of Isaac Newton} (Cambridge University Press, Cambridge).

\item[]
Norton J D (1984).  How Einstein Found His Field Equations: 
1912-1915, {\em Historical Studies in the Physical Sciences}, {\bf 
14}, 253--315.  Reprinted in {\it Einstein and the History of 
General Relativity: Einstein Studies}, D Howard and J Stachel 
eds., Vol.I, 101-159 ( Birkh\"auser, Boston).

\item[]
Penrose R (1995). {\em The Emperor's new mind} (Oxford University 
Press)

\item[]
Reisenberger M, Rovelli C (1997). Sum over Surfaces 
Form of Loop Quantum Gravity, {\it Physical Review}, {\bf D56}, 
3490--3508, gr-qc/9612035. 

\item[]
Rovelli C (1991a). What is observable in classical and quantum gravity?, 
{\em Classical and Quantum Gravity}, {\bf 8}, 297. 

\item[]
Rovelli C (1991b). Quantum reference systems, 
{\it Classical and Quantum Gravity}, {\bf 8}, 317.
  
\item[]
Rovelli C (1991c). Quantum mechanics without time: a model,
{\it Physical Review}, {\bf D42}, 2638. 

\item[]
Rovelli C (1991d). Time in quantum gravity: an hypothesis, 
{\it Physical Review}, {\bf D43}, 442.

\item[]
Rovelli C (1991e).  Quantum evolving constants. 
{\it Physical Review} {\bf D44}, 1339. 

\item[]
Rovelli C (1993a). Statistical mechanics of 
gravity and thermodynamical origin of time, {\it Classical and 
Quantum Gravity}, {\bf 10}, 1549.

\item[]
Rovelli C (1993b). The statistical state of the 
universe, {\it Classical and Quantum Gravity}, {\bf 10}, 1567.  

\item[]
Rovelli C (1993c). A generally covariant quantum field theory 
and a prediction on quantum measurements of geometry, 
{\it Nuclear Physics}, {\bf B405}, 797.

\item[]
Rovelli C (1995). Analysis of the different meaning of 
the concept of time in different physical theories, {\it Il Nuovo 
Cimento}, {\bf  110B}, 81.

\item[]
Rovelli C (1996). Relational Quantum Mechanics, {\it International 
Journal of Theoretical Physics}, {\bf 35}, 1637. 

\item[]
Rovelli C (1997a). Half way through the woods, in 
{\it The Cosmos of Science}, J Earman and JD Norton editors, 
(University of Pittsburgh Press and Universit\"ats Verlag 
Konstanz).

\item[]
Rovelli, C. (1997b) Loop Quantum Gravity,
Living Reviews in Relativity (refereed electronic journal),
http://www.livingreviews.org/ Articles/Volume1/1998-1rovelli; gr-qc/9709008

\item[]
Rovelli C (1998). Incerto tempore, incertisque loci: Can 
we compute the exact time at which the quantum measurement 
happens?, {\it Foundations of Physics}, {\bf 28}, 1031--1043, 
quant-ph/9802020.

\item[]
Rovelli C (1999). Strings, loops and the others: a critical 
survey on the present approaches to quantum gravity, in 
{\it Gravitation and Relativity: At the turn of the millennium}, N 
Dadhich J Narlikar eds (Poona University Press), gr-qc/9803024.

\item[]
Rovelli C and Smolin L (1988). Knot theory and quantum 
gravity, {\it Physical Review Letters}, {\bf 61}, 1155. 

\item[]
Rovelli C and Smolin L (1990). Loop space representation for quantum
general relativity, {\it Nuclear Physics}, {\bf B331}, 80.

\item[]
Rovelli C, Smolin L (1995a). Spin Networks and Quantum Gravity, 
{\it Physical Review}, {\bf D 53}, 5743.

\item[]
Rovelli C and Smolin L (1995b). Discreteness of Area and Volume 
in Quantum Gravity, {\em Nuclear Physics} {\bf B442}, 593.  
Erratum: {\em Nuclear Physics} {\bf B 456}, 734.  

\item[]
Smolin L (1997). The future of spin networks, gr-qc/9702030.  

\item[]
Stachel J (1989).  Einstein search for general covariance 
1912-1915, in {\em Einstein Studies}, D Howard and J Stachel eds, 
vol 1, 63-100 (Birkh\"auser, Boston).

\item[]
Stein H (1999). Physics and philosophy meet: the strange case of 
Poincar\'e. Unpublished.  

\item[]
Zeilinger A, in {\it Gravitation and Relativity: At the turn of 
the millennium}, N Dadhich J Narlikar eds (Poona University 
Press). 

\end{itemize}
 
\end{document}